\documentclass[onecolumn,superscriptaddress,nofootinbib]{revtex4-1}

\usepackage{amsmath,amssymb,caption,graphicx,epstopdf,mathtools,color,url}
\usepackage{physics}

\DeclareMathOperator{\erfc}{erfc}

\newcommand{\erfceqdim}{\erfc\Big(\sqrt{\dfrac{m}{2\overline{T}}}\dot{\overline{x}}\Big)}
\newcommand{\expeqdim}{\exp\Big(-\frac{m \dot{\overline{x}}^2}{2\overline{T}}\Big)}
\newcommand{\erfceqadim}{\erfc\Big( \frac{\epsilon\dot{x}}{\sqrt{2T}}\Big)}
\newcommand{\expeqadim}{\exp\Big(-\frac{\epsilon^2\dot{x}^2}{2T}  \Big)}

\newcommand{\deq}{\text{d.eq.}}

\renewcommand{\epsilon}{\varepsilon}
\newcommand{\ud}{\mathrm{d}}

\makeatletter 
\renewcommand{\pdv}[2]{\begingroup 
  \@tempswafalse\toks@={}\count@=\z@ 
  \@for\next:=#2\do 
    {\expandafter\check@var\next\@nil
     \advance\count@\der@exp 
     \if@tempswa 
       \toks@=\expandafter{\the\toks@}%
     \else 
       \@tempswatrue 
     \fi 
     \toks@=\expandafter{\the\expandafter\toks@\expandafter\partial\der@var}}%
  \frac{\partial\ifnum\count@=\@ne\else^{\number\count@}\fi#1}{\the\toks@}%
  \endgroup} 
\def\check@var{\@ifstar{\mult@var}{\one@var}} 
\def\mult@var#1#2\@nil{\def\der@var{#2^{#1}}\def\der@exp{#1}} 
\def\one@var#1\@nil{\def\der@var{#1}\chardef\der@exp\@ne} 
\makeatother 

\newcommand{\BigO}[1]{\mathcal{O}\left( #1 \right)}

\allowdisplaybreaks

\begin{document}

\title{Thermodynamics of slow solutions to the Gas-Piston equations}
\author{G. Gubbiotti}
\email{gubbiotti@mat.uniroma3.it}
\affiliation{ Universit\`a degli Studi Roma Tre, Dipartimento di Matematica e Fisica and Sezione INFN di Roma Tre}
\author{D. Chiuchi\`u}
\email{davide.chiuchiu@nipslab.org}
\affiliation{NiPS Lab, Universit\`a degli studi di Perugia, Dipartimento di Fisica e Geologia}

\begin{abstract}
Despite its historical importance, a perfect gas enclosed by a pistons and in contact with a thermal reservoirs is a system still largely under study. Its thermodynamic properties are not yet well understood when driven under non-equilibrium conditions. In particular, analytic formulas that describe the heat exchanged with the reservoir are rare. In this paper we prove a power series expansions for the heat when both the external force and the reservoir temperature are slowly varying over time but the overall process is not quasi-static. To do so, we use the dynamical equations from [Cerino \emph{et al.}, \textit{Phys. Rev. E}, \textbf{91} 032128] and an uncommon application of the regular perturbation technique.
\end{abstract}

\maketitle

\section{Introduction}
\label{sec:intro}

One of the main goals of non-equilibrium thermodynamics is to 
understand how a thermodynamic system evolves over time \cite{degrootmazur,seifert,sekimoto}.
The problem is difficult even in the simplest physical instances. {For example, the non-equilibrium behavior of the adiabatic piston \cite{gruber} is still a live research topic \cite[and references therein]{gislason,abreu,gruber}. Another problem which is conceptually simple, but difficult to treat is the non-equilibrium thermodynamics of 
a perfect gas enclosed by a cylindrical canister with a movable piston and 
in contact with a heat reservoir (see Figure \ref{fig:piston}). For this system there are multiple valid approaches: for example the one particle gas approach\cite{LuaGrosberg} and its legacy\cite{Baule,Hoppenau,Gong,Proesmans}, the explicit-friction formul\ae{} approach \cite{bizarro1,bizarro2,bizarro3}, and the gas particles-average approach  \cite{vulpiani1,Cerino2,Izumida,hisao}.
Among those references \cite{vulpiani1} is particularly interesting: there, the authors assumed that (i) the gas is perfect and 1-dimensional; (ii) the piston and each gas particle undergoes elastic collisions, so work is the energy exchanged in this way; (iii) the velocity of a gas particle is randomly changed according to the Maxwell-Boltzmann distribution of the reservoir when reservoir-gas particle collisions occurs \cite{tehver} and heat is the change in energy of the gas; (iv) the gas distribution is always Maxwellian although gas-reservoir and gas-piston collisions change the temperature of the gas over time. Combined in a laborious averaging process, the authors where able to derive a set of dynamical equations for the time evolution of the gas temperature $\overline{T}$ and piston position $\overline{x}$ according to any externally prescribed change of the external force $\overline{\Theta}$ and reservoir temperature $\overline{\Omega}$. 
In \cite{ChiuchiuGubbiotti01} we showed with the multiple scales method \cite{Kuzmak1959,KevorkianCole,Holmes,BenderOrszag}
and some technical assumptions, (discussed in Section \ref{sec:back})
that the equations derived in \cite{vulpiani1} allows 
to find an approximated expression for the heat exchanged with the reservoir
in two physically relevant cases: the \emph{relaxation to equilibrium}
and the \emph{slow isothermal compression}.
In the same paper \cite{ChiuchiuGubbiotti01} we pointed out the existence 
of particular solutions, which we called \emph{dynamical equilibrium solutions},
which describe the asymptotic behavior of the system when externally driven.

Stimulated by our previous understanding, we show in this paper
how to \emph{iteratively construct} the dynamical 
equilibrium solution at \emph{any desired precision order} through an uncommon application of the
regular perturbation theory \cite{KevorkianCole,Holmes,BenderOrszag}.
Our only assumptions will be that \emph{both}
the external force and the reservoir temperature 
are slow, smooth and non-vanishing.
Such solution immediately yields a general formula for 
the heat exchanged with the reservoir
as a formal power series in which \emph{all the coefficients}
are determined.
{Such kind of results are important due 
to their rarity in the literature and we believe that the method 
we used can be safely applied to other relevant problems in 
non-equilibrium thermodynamics.}
We suggest that similar power series approaches
can play a key role in the characterization of thermal cycles 
where drivings are slow but not
necessarily quasi-static. 

This paper is structured as follows: in Section \ref{sec:back}
we introduce the model from \cite{vulpiani1} together with 
the notations
and the formalism we will use throughout the paper;
in Section \ref{sec:dyneq} we derived our main results,
namely the full recursive relation from which we can
derive the coefficients of dynamical equilibrium solution
and the exchanged heat; in Section \ref{sec:disc} we comment
the thermodynamical relevance of our result, pointing also out 
similarities with the virial expansion, and
make some conjectures for possible future developments.

\begin{figure}[hbt]
\includegraphics[width=0.5\textwidth]{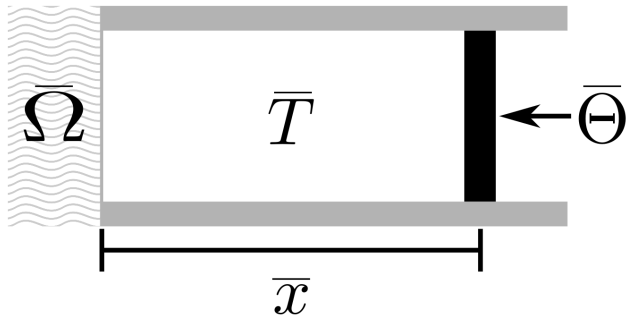}
\caption{{A schematic representation of the gas closed by a piston and in contact with a thermal reservoir. The quantities shown in the figure are the piston position $\overline{x}$, the gas temperature $\overline{T}$, the external force $\overline{\Omega}$ and the reservoir temperature $\overline{\Theta}$.}}\label{fig:piston}
\end{figure}

\section{Background}
\label{sec:back}

With the list of assumptions declared in Section \ref{sec:intro}, Cerino \emph{et. al.} were able to derive a set of equations to describe the dynamics of a gas inside a piston under the action of a variable external force when also the reservoir temperature changes over time. Such equations are
\begin{subequations}
\begin{align}
    &
    \begin{aligned}
\ddot{\overline{x}}&+\frac{\overline{\Theta}}{M}-\dfrac{\nu N}{\nu+1}\frac{1}{\overline{x}} \erfceqdim (\dot{\overline{x}}^2+\dfrac{\overline{T}}{m}) + \dfrac{\nu N}{\nu+1}\expeqdim\frac{\dot{\overline{x}}}{\overline{x}}\sqrt{\dfrac{2\overline{T}}{\pi m}}=0,\\
    \end{aligned}\label{eq:x_GPE}
    \\
    &
    \begin{aligned}
        \dot{\overline{T}}&+\frac{2\dot{\overline{x}}\left[m\dot{\overline{x}}^2+\overline{T}(1-2\nu)\right]}{\overline{x}(\nu+1)^2} \erfceqdim+\sqrt{\dfrac{2\overline{T}}{\pi m}}\frac{\overline{T}-\overline{\Omega}}{\overline{x}} +\frac{2}{\overline{x}(\nu+1)^2}\sqrt{\dfrac{2m\overline{T}}{\pi}}\left(\dfrac{2\nu\overline{T}}{m}-\dot{\overline{x}}^2\right)\expeqdim=0,
    \end{aligned}
\end{align}
    \label{eq:GPE}
\end{subequations}
where the upper dots denote time derivatives, $M$ 
is the mass of the piston, $m$ is the mass of a 
single gas molecule, $N$ is the total number of 
gas molecules, $\nu=m/M$, $\erfc(\cdot)$ is the 
complementary error function, $\overline{\Omega}$ is the externally controlled reservoir temperature, $\overline{\Theta}$ is the externally controlled force exerted on the piston, $\overline{x}$ is the piston position and $\overline{T}$ is the gas temperature. Eq.\eqref{eq:GPE} give a fairly good description of 
both the dynamics and the thermodynamics of the system \cite{vulpiani1}, the latter being expressed through 
\begin{subequations}
	\begin{align}
		\overline{E}&=M\frac{\dot{\overline{x}}^2}{2}+\overline{\Theta}\ \overline{x}+\frac{N									\overline{T}}{2},\\
		\overline{Q}(\overline{t}_i,\overline{t}_f)&=-\int_{\overline{t}_i}^{\overline{t}_f} \bigg[  		\overline{\Theta}\dot{\overline{x}}+ M\dot{\overline{x}}\ \ddot{\overline{x}} +\frac{N}{2}								\dot{\overline{T}}  \bigg]\, \mathrm{d}\overline{t}.
	\end{align}
    \label{eq:Energetics_dim}
\end{subequations}
where $\overline{E}$ is the total energy of the system \cite{jarzynski2} (composed by the kinetic energy of the piston, by the linear potential energy related to $\overline{\Theta}$ and by the thermal energy of the 1-Dimensional gas respectively), and $\overline{Q}$ is the heat \cite{seifert} that the system exchanges with the reservoir\footnote{Sign convention is such that $\overline{Q}>0$ for heat given to the reservoir.}. In \citep{ChiuchiuGubbiotti01} we showed that the thermodynamic limit
\begin{equation}\label{eq:thermodynamic_limits}
\frac{m}{M}=\nu\to0,\quad N\to\infty,\quad \epsilon= \sqrt{\frac{Nm}{M}}=\mbox{constant}
\end{equation}
can be taken in eq.\eqref{eq:GPE} if we first adimension them with the following transformation 
\begin{subequations}
\begin{align}
\overline{x}&=x\cdot \frac{\Omega_{r}N}{\Theta_r(1+\nu)g(\nu)}\\
\overline{T}&=T\cdot \frac{\Omega_{r}}{g(\nu)}\\
\overline{\Theta}&=\Theta\cdot \Theta_{r}\\
\overline{\Omega}&=\Omega\cdot \Omega_{r}\\
\overline{t}&=\frac{t}{\Theta_{r}}\cdot \sqrt{\frac{mN\Omega_{r}}{\nu(1+\nu)g(\nu)}} 
\end{align}
    \label{eq:dimensionless_variables}
\end{subequations}
where
\begin{equation}
g(\nu)=\frac{1+6\nu+\nu^2}{(1+\nu)^2}
\end{equation}
and $\Theta_r$ ($\Omega_{r}$) is an arbitrary force (temperature) reference value. Using these substitutions and then taking the thermodynamic limit eq.\eqref{eq:thermodynamic_limits} in eq.\eqref{eq:GPE}-\eqref{eq:Energetics_dim}, we obtain the following dimensionless dynamical equations
\begin{subequations}\label{eq:DGPE}
\begin{align}
    &\begin{aligned}
        \ddot{x}&+\Theta-\erfceqadim\frac{\epsilon^2\dot{x}^2+T}{x}
        +\expeqadim\frac{\dot{x}}{x}\sqrt{\dfrac{2T}{\pi}}\epsilon=0,
    \end{aligned}
    \\
    &\begin{aligned}
        \dot{T}&-2\erfceqadim \frac{\dot{x}}{x}(\epsilon^2\dot{x}^2+T)+\sqrt{\dfrac{2T}{\pi}}\frac{T-\Omega}{\epsilon x}-2\sqrt{\dfrac{2T}{\pi}}\epsilon\frac{\dot{x}^2}{x}\expeqadim=0.
    \end{aligned}
\end{align}
\end{subequations}
with the dimensionless $E$, and $Q$ densities:
\begin{subequations}
\begin{align}
E=&\frac{\overline{E}}{N\Omega_{r}}=\Theta x+\frac{\dot{x}^2}{2}+\frac{T}{2},\\
Q(\overline{t}_i,\overline{t}_f)=&\frac{\overline{Q}}{N\Omega_{r}}=-\int_{t_i}^{t_f} \bigg[  \Theta\dot{x}+ \dot{x}\ddot{x} +\frac{1}{2}\dot{T}  \bigg]\, \mathrm{d}t.\label{eq:Energetics_adim_Q}
\end{align}
    \label{eq:Energetics_adim}
\end{subequations}
The most important feature of eq.\eqref{eq:DGPE} is that they involve only the ratio $\epsilon$ of the total mass of the gas and the mass of the piston  as a physical parameter, a feature that this system shares with the adiabatic piston problem \cite{gruber}. Moreover, the gas has to be perfect, a condition that will likely end in having $\epsilon$ small. As a consequence, if a perturbation parameter is required for some approximated approach, $\epsilon$ is the most natural one to choose \cite{gruber}. An additional advantage of eq.\eqref{eq:DGPE} is that it's easy to check that their equilibrium condition is
\begin{equation}
    x_{\text{eq}}=\frac{\Omega}{\Theta}, \quad \dot{x}_{\text{eq}}=0, \quad T_{\text{eq}}=\Omega,
    \label{eq:equil}
\end{equation}
which is exactly what the perfect gas law and standard thermodynamics requires. If one is interested in situations where the system remains in the proximity of the equilibrium condition for $\Theta$ and $\Omega$, even if $\Theta$ and $\Omega$ changes over time, then eq.\eqref{eq:DGPE} can be linearized around eq.\eqref{eq:equil}. This process yields the following linear dimensionless equations
\begin{subequations}
\begin{align}
    &\begin{aligned}
        \ddot{x}&+\frac{\Theta^2}{\Omega}\bigg(x-\frac{\Omega}{\Theta}\bigg)
        +2\sqrt{\dfrac{2}{\pi \Omega}}\epsilon \Theta \dot{x}-\Theta \frac{T-\Omega}{\Omega}=0,
    \end{aligned}
    \label{eq:LDGPEa}
\\
&\dot{T}+2\Theta\dot{x}+\sqrt{\dfrac{2}{\pi \Omega}}\frac{\Theta}{\epsilon}(T-\Omega)=0
    \label{eq:LDGPEb}
\end{align}
\label{eq:LDGPE}
\end{subequations}

In \cite{ChiuchiuGubbiotti01} we assumed $\epsilon$ to be a ``small''
perturbative parameter,
the reservoir temperature 
to be constant (without loss of generality. $\Omega=1$)
and the external forcing $\Theta$ to be slowly varying over time
{(any function $B$ can be considered ``slow'' if 
$B=B(\epsilon t)$, which we assume from now on)}.
With such assumptions
we were able to find with the multiple scales method 
\emph{analytic approximated solutions} to \eqref{eq:LDGPE}. 
In particular we showed that
the relevant behavior of the system is described
by the following three scales:
\begin{equation}
    t_{0} =\frac{1}{\varepsilon}\int_{0}^{t} \Theta\left( \varepsilon \chi \right) \ud \chi,
    \quad
    t_{1} = \int_{0}^{t} \Theta\left( \varepsilon \chi \right) \ud \chi,
	\quad    
    t_{2} =\varepsilon t,
    \label{eq:scalest}
\end{equation}
where an explicit form of $\Theta$ 
it not required to give them an intuitive meaning.
The $t_{0}$-scale is the fastest one and
characterizes a transient suppression of the temperature of the gas.
From a physical point of view, it describes the indirect 
coupling of the piston with the reservoir.
The $t_{1}$-scale is the one at which transient oscillations of
the system are established. This reflects the direct coupling of the gas with the piston position and velocity. 
In the $t_{2}$-scale the exponential suppression
of the transient oscillatory terms appears and is the
proper scale of the external forcing too.

The approximated solution derived in \cite{ChiuchiuGubbiotti01}
 falls, as $t\to\infty$,
into a peculiar solution:
\begin{equation}
    x_{\deq}(t)
        =\frac{1}{\Theta\left( \varepsilon t \right)}
        + \varepsilon^{2} \left[ \frac{\Theta'\left(\varepsilon t\right)}{\Theta^{4}\left( \varepsilon t \right)}
            -2 \frac{(\Theta')^{2}\left(\varepsilon t\right)}{\Theta^{5}\left( \varepsilon t \right)}\right.
            \left.+ \left( 2\sqrt{\frac{2}{\pi}} + \sqrt{2\pi} \right)
            \frac{\Theta'\left( \varepsilon t \right)}{\Theta^{3}\left( \varepsilon t \right)} \right]
            +\BigO{\varepsilon^{4}}
    \label{eq:dynequil}
\end{equation}
with prime notation 
\begin{equation}
B'(s)=\left.\frac{\partial{B}(\mu)}{\partial \mu}\right|_{\mu=s}.
\end{equation}
We called this solution a \emph{dynamical equilibrium solution} as, from a physical point of view,  it describes
the system when it asymptotically follows the slow external driving. 
Mathematically this solution arise as the
particular solution of the inhomogeneous ordinary differential
equations coming from the multiple scales expansion and  can be computed in practice  
by taking all the integration 
constants in the multiple scale expansion equal to zero.
In \cite{ChiuchiuGubbiotti01}, we also hinted that the dynamical equilibrium solution 
\eqref{eq:dynequil} could have been obtained with the ``slow'' regular 
perturbation
\begin{equation}
x_\text{d.eq.}(t)=x_0(\epsilon t)+ \epsilon x_1(\epsilon t) +\epsilon^2 x_2(\epsilon t) + \epsilon^3 x_3(\epsilon t) +\BigO{\epsilon^4}.
\label{eq:ansatz}
\end{equation}
In the next Section we expand this point to show the generalization of the ansatz eq.\eqref{eq:ansatz} allows to \emph{iteratively construct the dynamical 
equilibrium solution at any order} when \emph{both}
the external force $\Theta$ and the reservoir temperature $\Omega$ are slow.

\section{Derivation of the dynamical equilibrium solution and of the heat}
\label{sec:dyneq}

Following the same procedure we used in \cite{ChiuchiuGubbiotti01} we
transform the system \eqref{eq:LDGPE} into a single third order equation.
At difference with \cite{ChiuchiuGubbiotti01}, where $\Omega\equiv1$,
we assume that
the external force and the reservoir temperatures are slow, 
i.e. $\Theta=\Theta(\epsilon t)$, $\Omega=\Omega(\epsilon t)$.
Then we solve \eqref{eq:LDGPEa} for $T$:
\begin{equation}
    T ={\frac {\Omega \left(\varepsilon t \right)  }{\Theta \left(\varepsilon t \right) }}\ddot{x}
    +2\sqrt{\frac{2}{\pi}}\sqrt{\Omega(\varepsilon t)}{\epsilon}\dot{x}
 +\Theta\left(\varepsilon t \right) x     
    \label{eq:subsT}
\end{equation}
and substitute in eq.
\eqref{eq:LDGPEb} to obtain:
\begin{widetext}
\begin{equation}
    \label{eq:x_iii}
    \begin{aligned}
        \frac{\Omega \left( \epsilon t \right)}{\Theta \left( \epsilon t \right) }\dddot{x}&+
        \left[ 2{\frac {\sqrt {\Omega \left( {\epsilon}t \right) }\sqrt {2}{
\epsilon}}{\sqrt {\pi }}}-{\frac {\Omega \left( {\epsilon}t
 \right)  \Theta'  \left( {\epsilon}t
 \right) {\epsilon}}{  \Theta^{2} \left( {\epsilon}t
 \right)}}+{\frac { \Omega' 
 \left( {\epsilon}t \right) {\epsilon}}{\Theta \left( {
\epsilon}t \right) }}+{\frac {\sqrt {\Omega \left( {\epsilon}t
 \right) }\sqrt {2}}{\sqrt {\pi }{\epsilon}}}\right]
        \ddot{x}\\         
        &+ \left[ 3\Theta \left( {\epsilon}t \right) +{\frac {\sqrt {2}{{
\epsilon}}^{2}\Omega'  \left( {\epsilon}t
 \right) }{\sqrt {\pi }\sqrt {\Omega \left( {\epsilon}t \right) }
}}+4{\frac {\Theta \left( {\epsilon}t \right) }{\pi }} \right]
        \dot{x}        
        + \left[  \Theta'   \left( {\epsilon}t \right) {
\epsilon}+{\frac {\sqrt {2} \left( \Theta \left( {\epsilon}t
 \right)  \right) ^{2}}{\sqrt {\pi }{\epsilon}\sqrt {\Omega
 \left( {\epsilon}t \right) }}}
 \right] x
        -{\frac {\sqrt {2}\Theta \left( {\epsilon}t \right) \sqrt {\Omega
 \left( {\epsilon}t \right) }}{\sqrt {\pi }{\epsilon}}}=0.
    \end{aligned}
\end{equation}
\end{widetext}

Now, instead of looking for a general approximation of $x(t)$
with the multiple scale method, we limit ourselves to the search for 
dynamical equilibrium solution $x_{\deq}(t)$. To this end
we must get rid of transient-like behaviors, i.e. those governed
by the $t_0$ and $t_1$ scales in \eqref{eq:scalest}.
The observation in \cite{ChiuchiuGubbiotti01} that the
dynamical equilibrium solution \eqref{eq:dynequil}
is obtainable from the ansatz \eqref{eq:ansatz} suggests
us to consider the following expansion:
\begin{equation}
x_{\deq}(t)=\sum_{n=0}^{\infty} x_n(\epsilon t)  \epsilon^n.
\label{eq:xseries}
\end{equation}
The series \eqref{eq:xseries} is a \emph{regular perturbation
expansion} with an uncommon feature: its coefficients
are functions of the slowest time scale $t_2$ in 
\eqref{eq:scalest}. With this choice all transient-like
behaviors are automatically ruled out, exactly as desired\footnote{This choice
is justified because it is a special case of the multiple scales method.
Therefore expansion validity and errors are ensured \emph{a priori}.}.
We then substitute \eqref{eq:xseries} in eq.\eqref{eq:x_iii}
and, upon defining 
\begin{equation}
s = \varepsilon t,\label{eq:s_subs}
\end{equation}
isolating $\epsilon$ coefficient, and then equating them 
to zero we obtain a system of \emph{linear algebraic equations for 
$x_n(s)$}. Solving it, we find:
\begin{subequations}
\begin{align}
    x_0(s)&= \frac{\Omega\left(s\right)}{\Theta\left(s\right)},
    \label{eq:x0}
    \\
    x_1(s)&=0,
    \label{eq:x1}
    \\
    x_2(s)&
    \begin{aligned}[t]
        &={\sqrt{\frac{2}{\pi}}\left(\pi+2\right) \frac {\Theta'  \left( s \right)\Omega^{3/2} \left( s \right)
	}{ \Theta^3 \left( s \right) }}-{\frac {\Omega''   \left( s \right) 
        \Omega \left( s \right) }{\Theta^{3}  \left( s \right) }}
        \\
        &+{\frac { \Omega^2  \left( s \right)  \Theta''  \left( s \right) }{  
	\Theta^4 \left( s \right)  }}-\frac{1}{2}\sqrt{\frac{2}{\pi}}\left(3\pi+4 \right){
            \frac {\sqrt {\Omega(s)} \Omega'\left( s \right) }{ \Theta^2  \left( s \right)  }}
        \\
        &+2{\frac {\Omega \left( s \right)  \Theta'   \left( s \right)  
	\Omega'   \left( s \right) }{ \Theta^4 \left( s \right)  }}
        -2{\frac {  \Omega^2 \left( s  \right) \left( \Theta' \right)^2  \left( s \right)  }{  
            \Theta^5 \left( s\right)  }},
    \end{aligned}
    \label{eq:x2}
    \\
    x_3(s)&=0,
    \label{eq:x3}
    \\
    x_n(s)&=L_2 x_{n-2}+L_4 x_{n-4}, \quad n\geq 4,
    \label{eq:xn}
\end{align}\label{eq:xterms}
\end{subequations}
where $L_2$ and $L_4$ are the linear differential operators:
\begin{subequations}
\begin{align}
L_2&=
-{\frac {\Omega \left( s \right) \left( s \right) }{\Theta^2 \left( s \right)  }} \dv[2]{s}
 	-\sqrt{\frac{1}{2\pi}}  {\frac {\sqrt {\Omega \left( s \right) } \left( 3\pi 
+4 \right) }{\Theta \left( s \right) }} \dv{s}
	-\sqrt{\frac{\pi}{2}} {\frac {
  \Theta' \left( s \right)    \sqrt {\Omega \left( s \right) }}{\Theta^2 \left( s \right)}}
\\
L_4&=
-\sqrt{\frac{\pi}{2}}{\frac {
  \Omega^{3/2} \left( s \right)}{\Theta^3 \left( s \right)}}\dv[3]{s}
-{\frac {  \Omega' \left( s \right)}{ \Theta^2 \left( s \right)}}\dv{s}
-\left[\frac{2\Omega\left( s\right)}{\Theta^2\left(s\right) } + \sqrt{\frac{\pi}{2}}\frac{\sqrt{\Omega\left(s\right)}}{\Theta^2\left(s\right)} \left(\frac{\Omega}{\Theta} \right)' \left(s\right) \right]
 .
\end{align}\label{eq:Loperators}
\end{subequations}
Eq.\eqref{eq:xn} defines a recurrence relation where
 $x_{0}(s)$, $x_{1}(s)$, $x_{2}(s)$ and $x_{3}(s)$
act as initial conditions. We can therefore use eq.\eqref{eq:xterms} and eq.\eqref{eq:Loperators} to obtain
every term in the series \eqref{eq:xseries} simply by differentiation
and multiplication. Furthermore, eq.\eqref{eq:xn} is homogeneous and $x_{1}=x_{3}=0$, then $x_{2n+1}=0$ for every $n\in \mathbb{N}$.
A sample plot of $x_\deq(t)$ and the corresponding $T_\deq(t)$ with
\begin{equation}
\Omega(s)=1+\frac{1}{10}\sin(s), \quad \Theta(s)=1+\frac{1}{10}\cos(s)
\label{eq:thetaomega}
\end{equation}
is shown in Figure\eqref{fig:xT}.

\begin{figure}[t]
	\centering
	\includegraphics[width=0.5\textwidth]{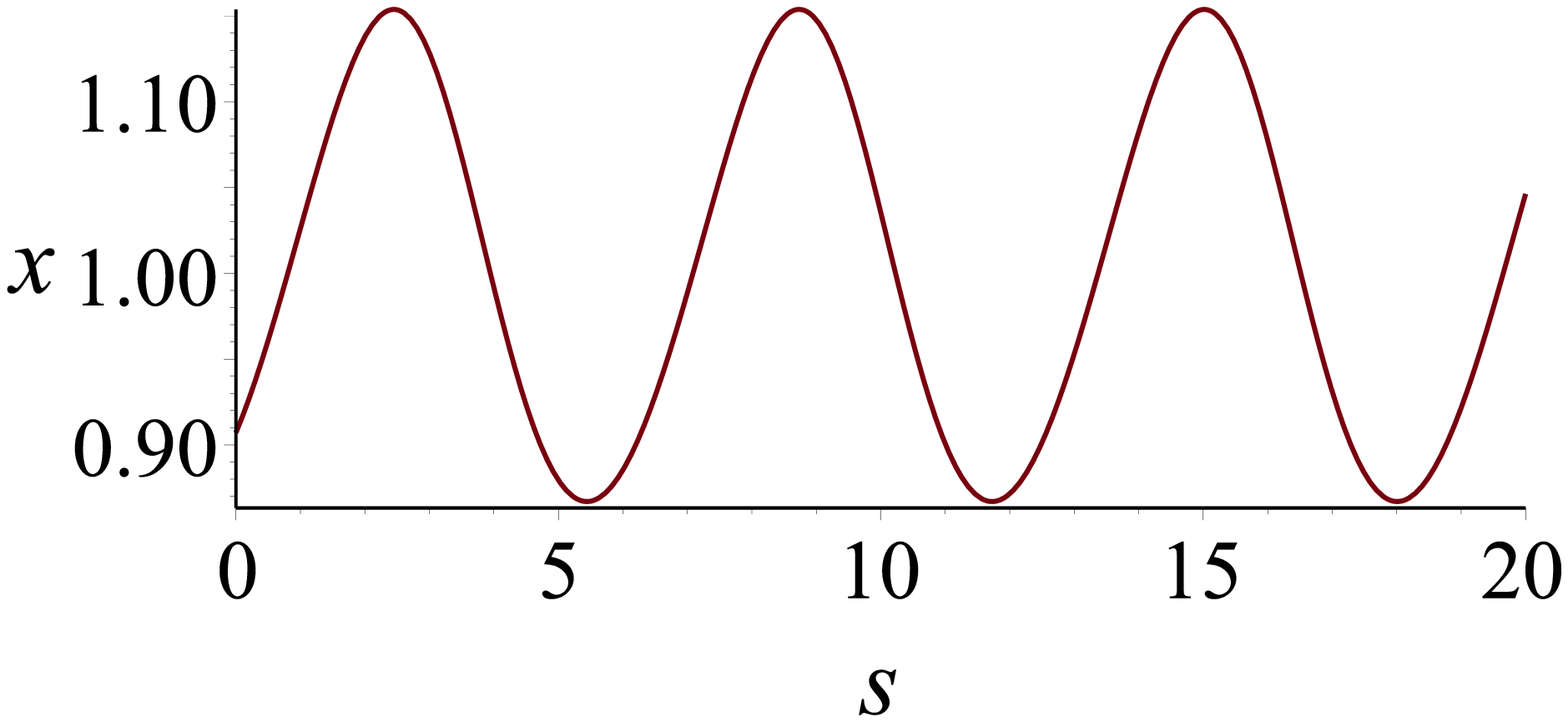}\includegraphics[width=0.5\textwidth]{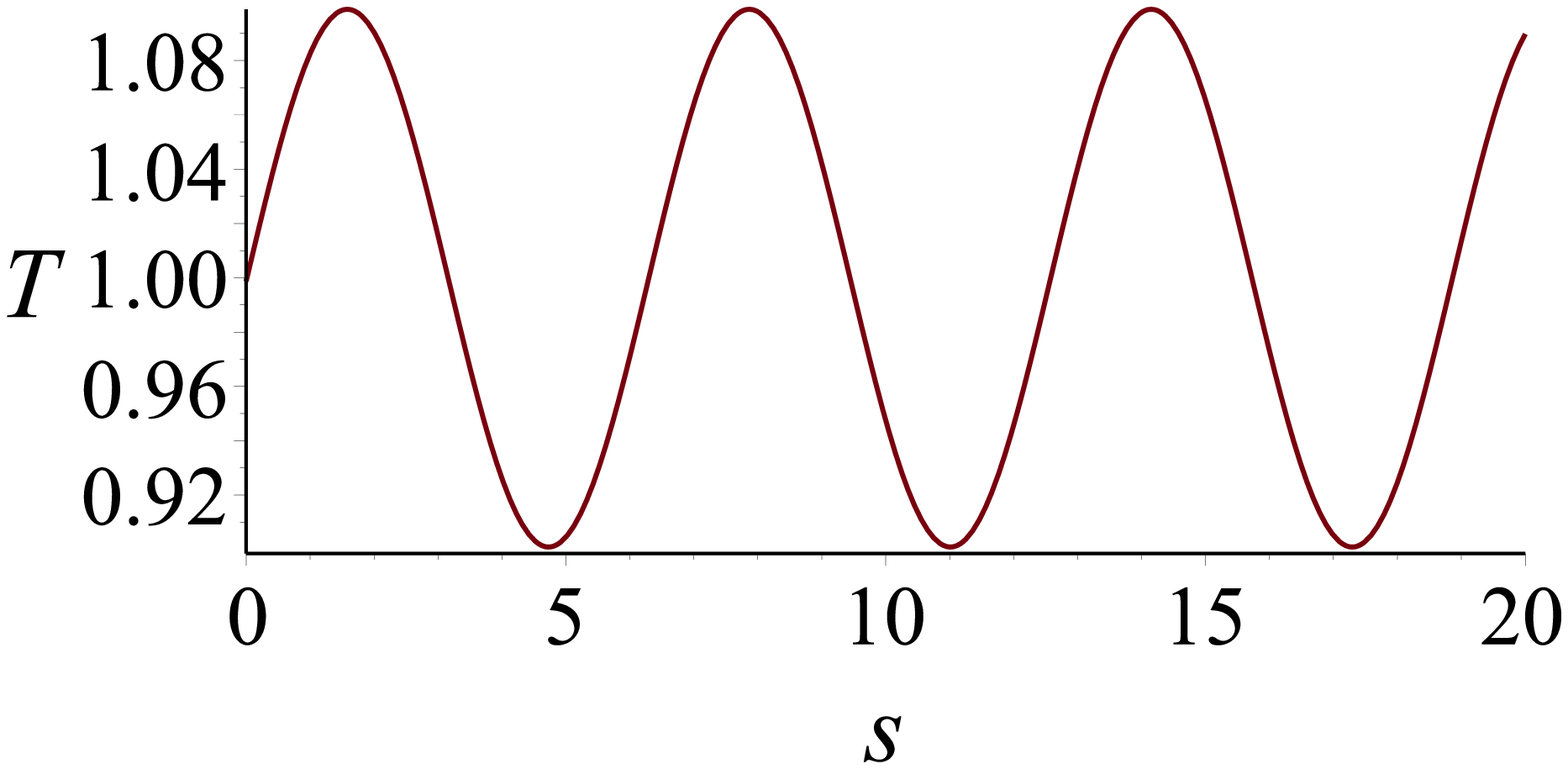}
	\caption{{Sample trajectories for $\epsilon=1/16$ with the first three non-zero terms of $x$ and $T$ expansions for ${\Omega(s)=1+0.1\sin(s)}$ and $\Theta(s)=1+0.1\cos(s)$.}}\label{fig:xT}
\end{figure}

Knowing the terms of the expansion
\eqref{eq:xseries}, we can find an analogous series
expansion for the the heat:
\begin{equation}
    Q(0,s)=\sum_{n=0}^{\infty} Q_{2n}(0,s)\epsilon^{2n}.
    \label{eq:heatseries}
\end{equation}
Indeed, substituting eq.\eqref{eq:s_subs} 
into \eqref{eq:Energetics_adim_Q} we
first find:
\begin{equation}
\begin{aligned}
    Q(0,s)=&-\int_{0}^{s} \Theta(s) x'(\sigma) \ud\sigma-\frac{\epsilon^2}{2}\left[x'(s)-x'(0) \right]\\ 
    &- \frac{T(s)-T(0)}{2}
\end{aligned}
\end{equation}
which, upon the substitution of \eqref{eq:xseries} and \eqref{eq:subsT}, 
yields the following values for the coefficients in 
\eqref{eq:heatseries}:
\begin{subequations}
\begin{align}
    Q_0(0,s)&=\Theta \left( s \right) \left[x_0\left( 0\right) -x_0\left(s\right) \right]
    -\int_{0}^{s}\!\Theta \left( \sigma \right)   x_0'  \left( \sigma \right) \ud\sigma
    \\
    Q_{2n}(0,s)&
    \begin{aligned}[t]
        =&-\int_0^{s} \Theta\left( \sigma\right) x_{2n}'(\sigma)\ud\sigma
        -\frac{1}{2}\sum_{l=0}^{n-1} \left[x'_{2l}(s)x'_{2n-2l-2}(s)- x'_{2l}(0)x'_{2n-2l-2}(0) \right]
        -\left[\Theta(s)x_{2n}(s)-\Theta(0)x_{2n}(0) \right]
        \\
        &-2\sqrt{\frac{2}{\pi}}\left[\sqrt{\Omega(s)}x'_{2n-2}(s)-\sqrt{\Omega(0)}x'_{2n-2}(0) \right]
        - \left[\frac{\Omega}{\Theta}(s)x''_{2n-2}(s)-\frac{\Omega}{\Theta}(0)x''_{2n-2}(0) \right].
    \end{aligned}
\end{align}
\label{eq:qn}
\end{subequations}
It is worth to note that all the coefficients \eqref{eq:qn}
 can, in principle, be determined from the recurrence
relation for $x_n$ \eqref{eq:xterms}. An integration is
involved, which may be non-trivial, but, from a formal
point of view, the series is well defined for sufficiently small values of $\epsilon$.

\section{Discussion}
\label{sec:disc}

In this concluding section, we comment the relevance of eq.\eqref{eq:xterms} and of eq.\eqref{eq:qn} and point out interesting features. Speaking specifically of the dynamical equilibrium solution, eq.\eqref{eq:xterms} possess remarkable properties. The first one is that,it allows to calculate the dynamical equilibrium solution $x_\deq$
and hence $T$ via \eqref{eq:subsT} with an arbitrary precision.
Only multiplications and differentiations are required in such
formula, thus making the evaluation of $x$ and $T$ a trivial task.
For practical purposes, the evaluation of the series
\eqref{eq:xseries} must be truncated at some point.
Being absent the
odd terms, if we stop at the $N$th iteration we will have an error
of order $\BigO{\varepsilon^{2N+2}}$, which gives us a pretty high
precision with few iterations.
Obviously this is valid only within
the convergence radius; this can be in principle determined
from the recurrence relation \eqref{eq:xn}, but 
its determination is a complex task far beyond the scope of this paper.

The second remarkable property is that 
\eqref{eq:xn} is valid for general instances of
$\Theta$ and $\Omega$ provided, that they (i) depend on time only through $\epsilon t$, (ii) they are smooth and (iii) never
vanish.
This last condition must necessary hold for $\Theta$,
while it can be slightly relaxed for $\Omega$.
\emph{A priori} the vanishing of both $\Theta$ and $\Omega$ 
introduces singularities in eq.\eqref{eq:x_iii}. 
As eqs.(\ref{eq:xterms},\ref{eq:Loperators}) show, we can't get rid 
of the singularities introduced by $\Theta$. 
Conversely from the same equations we see that the singularities in
$\Omega$ do not explicitly appear in the solution, so they
are in a certain sense \emph{removable}. 
However the price to pay is that the solution is compelled to live 
in a compact subset of the real line:
if there exists a $s_0>0$ such that $\Omega(s_0)=0$ then 
our solution is defined only on the subset $0\leq s<s_0$
due to the presence of square roots in eqs.(\ref{eq:xterms},\ref{eq:Loperators}). 
Physically $\Omega(s_0)=0$ means that we allow the reservoir temperature 
to reach the absolute zero in a finite time, which is forbidden 
by thermodynamics as it means that all the degrees of freedom of 
the system becomes frozen. As a consequence a vanishing
$\Omega(s)$ is mathematically acceptable (even if with some
limitations), but clearly not physically.

The third interesting feature   of eq.\eqref{eq:xseries} is
that it extends the results of our previous
paper \cite{ChiuchiuGubbiotti01} where we were limited
to use $\Omega=1$ and opens the possibility to study thermodynamic
cycles. In turn, using this regular expansion technique,
we lost the characterization of transient phenomena.
This is not a major drawback, if for example one
is interested in the study of the thermodynamic cycles
efficiency, where transient behaviors are
neglected. Moreover numerical evidence lead
us to conjecture that, for $\varepsilon$ sufficiently
small, all solutions of \eqref{eq:LDGPE} will at some time fall into our
dynamical equilibrium solution. 
One example of such evidence is obtained as follows: using
the particular form of $\Theta$ and $\Omega$ given in
eq.\eqref{eq:thetaomega}
we simulate eq.\eqref{eq:x_iii} with $x(0)=x_{\deq}(0)$ 
and $x(0)=x_{\deq}(0)+K\epsilon$, thus obtaining 
$x^{\text{num}}(t)$ and $x^{\text{num}}_{\deq}(t)$ 
respectively. We then study 
\begin{equation}
R=\frac{\displaystyle\max_{s\in[7\pi, 9\pi]}\left| x^\text{num}(s)-x_{\deq}(s) \right|}{\displaystyle\max_{s\in[7\pi, 9\pi]}\left| x^\text{num}_{\deq}(s)-x_{\deq}(s) \right|}
\end{equation}
as a function of $K$ and $\epsilon$, where $x_\deq$ is truncated at $n=4$ and $s\in[7\pi,9\pi]$ ensures that we are way ahead of the time evolution of the system. As it emerges by Figure \ref{fig:attractor}, $R\approx 1$ for various value of $N$ and $\epsilon$, so , we reasonably say that $x^\text{num}(s)$ is attracted by eq.\eqref{eq:xseries}. This means that the dynamical equilibrium solution has a quite general usefulness.

\begin{figure}[hbt]
    \centering
    \includegraphics[width=0.6\textwidth]{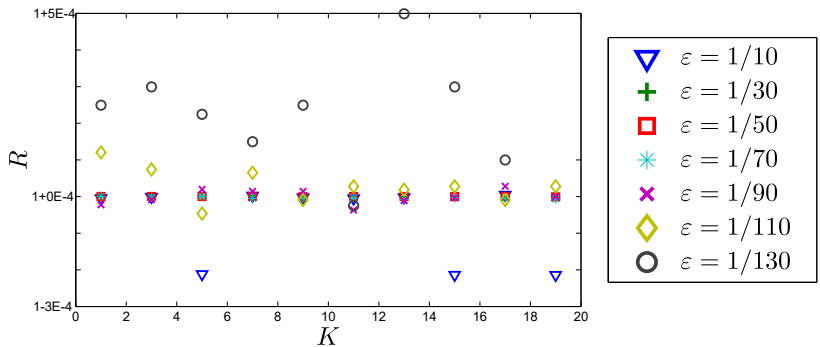}
    \caption{Plot of $R$ as a function of $K$ for different values of $\epsilon$.}
    \label{fig:attractor}
\end{figure}

The last interesting feature of the expansion \eqref{eq:xseries} 
is that if we limit ourselves
to the $0$-th order term, we obtain the law of perfect gasses. 
Higher orders terms are small corrections arising from the fact that 
$\Omega$ and $\Theta$ are changed in a finite time. 
In this sense  eq.\eqref{eq:xseries} and eq.\eqref{eq:xterms} 
can be qualitatively considered as an analogue of the virial 
expansion for a perfect gas where finite time-effects plays the 
role of inter-particle interaction contributions. Clearly, this analogy is a formal one: in \cite{vulpiani1} the gas is assumed to be perfect and remains perfect during the whole time evolution. However it is not unreasonable to think that the corresponding equation will have as $0$-th order term the usual virial theorem for real gasses if the gas particles are allowed to interact. This suggests the existence of higher order time-dependent virial theorems.

Coming to the relevance of \eqref{eq:heatseries} and \eqref{eq:qn}, they 
allow us to calculate the heat as a formal series in which 
all the coefficients are well determined through
a purely mechanical model \cite{vulpiani1}.
Eq.\eqref{eq:qn} is
therefore a quite uncommon result in the framework of
non-equilibrium thermodynamics, as similar formal results are rare in
the literatur. It gives an explicit expression
of heat exchanged at any intermediate time that can be
evaluated \emph{a priori} with the sole knowledge of
the driving protocol. This opens interesting scenarios to analyze.
For example it reduces the evaluation of heat in practice
to the problem of finding the terms in \eqref{eq:xseries}
and to performing the integration in eq.\eqref{eq:qn}.
Another interesting possibility is the formal evaluation of the efficiency
of thermodynamic cycles when drivings are slow but not 
necessarily quasi-static, i.e. to characterize situations where 
efficiency is expected to be high while the power is close to zero.
{We point out that in the original model we are not allowed to consider adiabatic transformations. Therefore we are unable to describe Carnot cycles but it is possible to study Ericsson-like cycles\cite{vulpiani1,Cerino2}, with the only caveat that $\Omega$ and $\Theta$ must be smooth function. This may seems an internal contradiction as Ericsson cycle is not smooth; however, any periodic functions can be reasonably approximated by a smooth trigonometric polynomial\cite{Whittaker}. To satisfy the ``slow'' driving condition such trigonometric polynomial should not contain high order harmonics, but this is usually sufficient to yield a good approximation.} 


\begin{acknowledgments}
DC is supported by the 
European union (FPVII(2007-2013) under G.A. n.318287 LANDAUER).
GG is supported by  INFN  IS-CSN4 \emph{Mathematical Methods of
Nonlinear Physics}.
\end{acknowledgments}

\bibliographystyle{unsrt}
\bibliography{references_davide}

\end{document}